\documentclass[pdflatex,sn-mathphys-num,a4paper,oneside]{sn-jnl}% Math and Physical 

\usepackage{graphicx}%
\usepackage{multirow}%
\usepackage{amsmath,amssymb,amsfonts}%
\usepackage{amsthm}%
\usepackage[title]{appendix}%
\usepackage{xcolor}%
\usepackage{textcomp}%
\usepackage{manyfoot}%
\usepackage{booktabs}%
\usepackage{algorithm}%
\usepackage{algorithmicx}%
\usepackage{algpseudocode}%
\usepackage{listings}%
\usepackage{enumerate}
\usepackage{siunitx}
\usepackage{fontawesome5} % Font Awesome 5
\usepackage{wrapfig}
\usepackage{scalerel}
\usepackage{lipsum}
\usepackage{marginnote} % more robust than \marginpar

\usepackage{geometry}

\makeatletter
\pretocmd{\@authoraddress}{\aftergroup\newpage}{}{}
\makeatother

\makeatletter
\renewcommand{\addressfont}{%
  \reset@font\fontsize{8bp}{10bp}\selectfont\titraggedcenter%
}
\makeatother

\begin{document}

\title[Sustainable Computational Science and Engineering]{\vspace{-2cm}Perspectives on Sustainable Computational Science and Engineering}

% how to name authors

\author*[1]{\fnm{Julia} \sur{Kowalski}}\email{kowalski@mbd.rwth-aachen.de}
\equalcont{These authors contributed equally to the writing of this manuscript.}

\author[2]{\fnm{Ann S.} \sur{Almgren}}
\equalcont{These authors contributed equally to the writing of this manuscript.}
\author[3]{\fnm{Achim} \sur{Basermann}}
\equalcont{These authors contributed equally to the writing of this manuscript.}
\author[1]{\fnm{Alan} \sur{Correa}}
\equalcont{These authors contributed equally to the writing of this manuscript.}
\author[3]{\fnm{Immo} \sur{Huismann}}
\equalcont{These authors contributed equally to the writing of this manuscript.}
\author[1]{\fnm{Uwe} \sur{Naumann}}
\equalcont{These authors contributed equally to the writing of this manuscript.}
\author[4]{\fnm{Hendrik} \sur{Ranocha}}
\equalcont{These authors contributed equally to the writing of this manuscript.}
\author[5]{\fnm{Matthias} \sur{Rauter}}
\equalcont{These authors contributed equally to the writing of this manuscript.}
\author[6]{\fnm{Jens} \sur{Saak}}
\equalcont{These authors contributed equally to the writing of this manuscript.}
\author[7]{\fnm{Michael} \sur{Schlottke-Lakemper}}
\equalcont{These authors contributed equally to the writing of this manuscript.}
\author[8]{\fnm{Robert} \sur{Speck}}
\equalcont{These authors contributed equally to the writing of this manuscript.}
\author[1]{\fnm{Manuel} \sur{Torrilhon}}
\equalcont{These authors contributed equally to the writing of this manuscript.}
\author[9]{\fnm{Mario} \sur{Wolter}}
\equalcont{These authors contributed equally to the writing of this manuscript.}

\author[1]{\fnm{Niklas} \spfx{von der} \sur{Assen}}
\author[8]{\fnm{Maria Guadalupe} \sur{Barrios Sazo}}
\author[1]{\fnm{Dipankul} \sur{Bhattacharya}}
\author[10]{\fnm{Marcel} \sur{Bock}}
\author[1]{\fnm{Faras} \sur{Brumand-Poor}}
\author[8]{\fnm{Dirk} \sur{Brömmel}}
\author[1]{\fnm{Semih} \sur{Burak}}
\author[11]{\fnm{Sara} \sur{Grundel}}
\author[12]{\fnm{Christiane} \sur{Helzel}}
\author[13]{\fnm{Bruce} \sur{Hendrickson}}
\author[14]{\fnm{Robert} \sur{Klöfkorn}}
\author[15]{\fnm{Katharina} \sur{Kormann}}
\author[16]{\fnm{Zahra} \sur{Lakdawala}}
\author[1,8]{\fnm{Hannah} \sur{Lanzrath}}
\author[1]{\fnm{Camilla} \sur{Lüttgens}}
\author[1]{\fnm{Anna} \sur{Matuszyńska}}
\author[1]{\fnm{Matthias} \sur{Meinke}}
%\author[1]{\fnm{Simon} \sur{Märtens}}
\author[1,8]{\fnm{Abigail} \sur{Morrison}}
\author[17]{\fnm{William S.} \sur{Moses}}
\author[1]{\fnm{Georgii} \sur{Oblapenko}}
%\author[13]{\fnm{Tapasya} \sur{Patki}}
\author[18]{\fnm{Marcel} \sur{Pfeiffer}}
\author[1,19]{\fnm{Jana} \sur{Sasse}}
\author[1]{\fnm{Katharina} \sur{Schmitz}}
\author[18]{\fnm{Miriam} \sur{Schulte}}
\author[18]{\fnm{Anna} \sur{Schwarz}}
\author[1]{\fnm{Pit} \sur{Steinbach}}
\author[1]{\fnm{Ingo} \sur{Steldermann}}
\author[1,8]{\fnm{Anton} \sur{Stratmann}}
\author[1]{\fnm{Lambert} \sur{Theisen}}
\author[20]{\fnm{Joseph} \spfx{De} \sur{Veaugh-Geiss}}
\author[1]{\fnm{Tim} \sur{Wegmann}}
\author[21]{\fnm{Gabriel} \sur{Wittum}}
\author[1]{\fnm{Anil} \sur{Yildiz}}

\affil[1]{\orgname{RWTH Aachen University}, \city{Aachen}, \country{Germany}}
\affil[2]{\orgname{Lawrence Berkeley National Laboratory}, \city{Berkeley}, \country{USA}}
\affil[3]{\orgname{German Aerospace Center (DLR)}, \country{Germany}}
\affil[4]{\orgname{Johannes Gutenberg University Mainz}, \city{Mainz}, \country{Germany}}
\affil[5]{\orgname{The Qt Company}, \city{Oslo}, \country{Norway}}
\affil[6]{\orgname{Max Planck Institute for Dynamics of Complex Technical Systems}, \city{Magdeburg}, \country{Germany}}
\affil[7]{\orgname{University of Augsburg}, \city{Augsburg}, \country{Germany}}
\affil[8]{\orgname{Forschungszentrum Jülich GmbH}, \city{Jülich}, \country{Germany}}
\affil[9]{\orgname{Technische Universit\"at Braunschweig}, \city{Braunschweig}, \country{Germany}}
\affil[10]{\orgname{University of Oldenburg}, \city{Oldenburg}, \country{Germany}}
\affil[11]{\orgname{Leipzig University of Applied Sciences}, \city{Leipzig}, \country{Germany}}
\affil[12]{\orgname{Heinrich Heine University Düsseldorf}, \city{Düsseldorf}, \country{Germany}}
\affil[13]{\orgname{Lawrence Livermore National Laboratory}, \city{Livermore}, \country{USA}}
\affil[14]{\orgname{Lund University}, \city{Lund}, \country{Sweden}}
\affil[15]{\orgname{Ruhr University Bochum}, \city{Bochum}, \country{Germany}}
\affil[16]{\orgname{Fraunhofer Institute for Wind Energy Systems}, \city{Oldenburg}, \country{Germany}}
\affil[17]{\orgname{University of Illinois Urbana-Champaign}, \city{Champaign}, \country{USA}}
\affil[18]{\orgname{University of Stuttgart}, \city{Stuttgart}, \country{Germany}}
\affil[19]{\orgname{TU Dortmund University}, \city{Dortmund}, \country{Germany}}
\affil[20]{\orgname{KDE e.V.}, \city{Berlin}, \country{Germany}}
\affil[21]{\orgname{King Abdullah University of Science and Technology}, \city{Thuwal}, \country{Saudi Arabia}}

%%==================================%%
%% Abstract                         %%
%%==================================%%

\abstract{Computational Science and Engineering (CSE) combines expertise at the intersection of engineering, applied mathematics, and computer science to form powerful methods for model-based design and model-based decision support across disciplines. Today, CSE methods and tools have an impact as an enabling technology in the development of increasingly sustainable products, processes, and operations. However, sustainability is rarely considered holistically in the context of CSE itself. This paper develops a perspective on sustainability in CSE by distinguishing CSE as an enabler of sustainability in application domains from sustainability within CSE itself, which rests on two complementary pillars: sustainable computing and sustainable software. We present illustrative examples for these pillars and derive recommendations and best practices for CSE stakeholders. Developing a comprehensive understanding of how sustainability can create added value in CSE is an important future direction for the field. This calls for a concerted effort within the CSE community to define measurable outcomes and best practices that embed sustainability as a core design principle rather than an afterthought.}




\maketitle

%\newpage
\section{Introduction}
\label{sec:introduction}

Computational Science and Engineering (CSE) combines expertise at the intersection of engineering, applied mathematics, and computer science to form powerful methods for model-based design and model-based decision support across multiple disciplines. The field of CSE facilitates the quantified retrospection, introspection, and prediction of multiphysics and multiscale processes,  such as fluid flow, deforming and transforming matter, propagating heat, radiation in heterogeneous media, and phase-transition processes. Those processes are relevant in the design of aircraft and turbomachinery, wind-energy systems, combustion and energy-conversion devices, batteries and fuel cells, climate and weather prediction, manufacturing -- to name just a few. Research directions in CSE range from the traditional development of numerical methods for error-controlled modeling of multiphysics and multiscale processes and algorithms for high-performance systems to topics that have gained recent attention
such as Bayesian methods, model reduction \cite{morBenGQetal21,morBenGQetal21a,morBenGQetal21b}, scientific machine learning \cite{rackauckas2020ude_sciml}, and foundations of digital twins \cite{nationalacademies2024digitaltwins}. Methodological and application-focused research in CSE acts as a driver of innovation in all fields of science and engineering, an example being the reverse mode of algorithmic differentiation, which by
now is a key ingredient for parameter optimization in modern neural networks~\cite{griewank2012reverse_mode}. 

In recent years, artificial intelligence (AI) and machine learning (ML) have considerably expanded the methodological scope of CSE, opening new avenues for scientific discovery and engineering design. The greatest advances are expected not from AI replacing computational science, but from a principled integration of the two. Rather than replacing established computational methods, AI creates new opportunities to combine data-driven learning with decades of advances in mathematical modeling, numerical analysis, and scientific computing. This convergence has given rise to scientific machine learning and physics-informed or physics-guided artificial intelligence, where learning algorithms are designed to exploit physical laws, conservation principles, and domain knowledge alongside observational data. Such hybrid approaches may enable computational models that are not only more data-efficient and reliable, but also capable of generalizing beyond the conditions represented in the training data. They enable highly efficient surrogate models for complex simulations, accelerate numerical solution procedures, and provide integral components for digital twins and computational decision-support systems \cite{nationalacademies2024digitaltwins}.

More broadly, the growing methodological diversity of CSE has led to increasingly complex computational workflows. This trend predates AI but has been significantly accelerated by its emergence. Consequently, resource-aware methodologies developed within CSE are becoming essential not only for improving the efficiency of individual algorithms and software components, but also for optimizing complete end-to-end computational workflows. Such a holistic perspective is indispensable for ensuring that future computational ecosystems deliver scientific insight at acceptable economic and environmental cost.

Energy consumption and environmental impact are therefore receiving increasing attention~\cite{lannelongue2023greener,kovalchuk2025computational}. These concerns are commonly viewed as key aspects of sustainability, often associated with preserving natural resources and maintaining ecological balance. More broadly, the United Nations defines sustainable development as development that \emph{``meets the needs of the present without compromising the ability of future generations to meet their own needs''}~\cite{WCED1987}. Although intentionally broad, this definition is commonly interpreted through three interconnected dimensions: environmental, economic, and social sustainability. Over the past decade, it has become increasingly evident that sustainability is not only a societal objective but also a guiding principle for scientific research. It shapes both the research questions being pursued and the way research itself is conducted. Translated to CSE, it requires computational methodologies that are not only scientifically sound but also resource-efficient, reproducible, and sustainable.

An operational challenge inherent in sustainability is the separation of timescales between decisions and the systems they affect. Computational and development choices are often made on the timescales of individual projects, whereas software maintenance needs, scientific capabilities, infrastructure impacts, and the benefits enabled in application domains unfold over much longer hardware and software related life cycles. Sustainability in CSE therefore cannot be assessed solely through a single immediate objective such as runtime, accuracy, or energy consumption. It requires multi-objective, life-cycle-aware assessment that relates short-term computational and development decisions to longer-term scientific value. Sustainability as a research goal has been prominent in projects that leverage CSE methods as enabling technologies to optimize comprehensively in high-dimensional design spaces, propagate uncertainties, and conduct parameter estimation. In this context, CSE creates added value in the design of sustainable products and in policy making \cite{baker2018brn_sciml}.
%In addition, applied sciences and engineering fields benefit from recent improvements in digital twin technologies \cite{nationalacademies2024digitaltwins} that facilitate design, decision support, and diagnostics in model-based engineering workflows. 
Yet, while CSE has become a powerful enabler of sustainability across many scientific and engineering disciplines, comparatively little attention has been devoted to the sustainability of CSE itself, and the relevant dimensions are rarely considered in an integrated and holistic manner. Addressing this gap is the goal of this paper.

We start with a discussion of how CSE can act as an enabler of research for sustainable development, featuring two examples in Section~\ref{sec:enabler}. Section~\ref{sec:toward_sustainable_cse} introduces our framework for sustainability within CSE. Sections \ref{sec:sustainable_computing} and \ref{sec:sustainable_software} then present its two operational pillars—sustainable computing and sustainable software—through illustrative examples, and a discussion of how these two sustainability perspectives can be integrated into future-directed CSE research. We conclude that fostering this requires action by all stakeholders, and we derive recommendations for CSE stakeholders in Section~\ref{sec:recommendations}. Finally, we close with a summary and outlook in Section~\ref{sec:outlook}.  

\section{CSE as Sustainability Enabler}
\label{sec:enabler}

The following two examples illustrate selected ways in which CSE can act as an enabling factor for sustainability research across a wide range of application domains.

\subsection{Designing Next-Generation Aircraft}\label{subsec:aircraft}

To reduce the environmental impact of aviation, the European Commission articulated long-term research and innovation goals in its Flightpath 2050 vision~\cite{eu_2011_flightpath2050}. Relative to the capabilities of a typical new aircraft in 2000, the vision envisages technologies and procedures available by 2050 that enable a 75\% reduction in CO$_2$ emissions per passenger-kilometre, a 90\% reduction in nitrogen oxide emissions, and a 65\% reduction in the perceived noise emission of flying aircraft. This strategy is not unique to Europe, and similar goals are promoted in NASA's Environmentally Responsible Aviation~\cite{collier_2010_nasa_era}. Achieving these goals will require completely new aircraft designs. At the same time, even small improvements in aerodynamic efficiency, propulsion efficiency, or operating performance can translate into substantial CO$_2$ reductions when accumulated over large fleets and long service lifetimes. This makes aircraft design a particularly relevant example of how CSE can enable sustainability.

Meeting these goals requires novel methods capable of systematic exploration of large multidisciplinary design spaces \cite{slotnick_2014_cfd2030}. The relevant design decisions concern aerodynamics, structures, propulsion, fuel choice, acoustics, and operations. These interdependencies can only be reliably assessed with advanced modeling, simulation, and optimization. CSE is therefore a central enabler of model-based aircraft design. Examples of relevant design objectives are:

\begin{itemize}
  \item overall airframe geometry~\cite{bezos_2011_fuel_efficiency_through_airframe_design}, whether through a slight change in wing design or a transition from a traditional to a blended wing-body configuration for noise reduction~\cite{collier_2010_nasa_era};
  \item the propulsion system, e.g.,\ moving to an open rotor design~\cite{hughes_2011_green_engine_technology}, improving the turbine's fuel efficiency, moving towards greener fuels at the expense of lowered fuel efficiency per kilogram~\cite{boretti_2024_green_fuels}, or a transition to a partially or fully electric propulsion system~\cite{fard_2022_electric_flight,epstein_2019_electric_flight}.
\end{itemize}

These changes result in aircraft designs that deviate from the well-established design space of current aircraft. Even when restricting the design space to a traditional tube-and-wing configuration, gradient-based optimization using computational fluid dynamics (CFD) for external aerodynamics and computational structural mechanics (CSM) for structural analysis yielded a~$3.6\%$ reduction in fuel consumption~\cite{goertz_2017_mdo}. The gains can be increased further as the design space addressed during optimization grows and more design alternatives are considered. For example, a NASA assessment of a hybrid wing-body configuration found that integrated airframe and propulsion technologies could achieve the project's goal of 42~dB cumulative below Stage~4 certification limits~\cite{collier_2010_nasa_era}.

All of these approaches rely heavily on capabilities developed in CSE, including flow simulations ranging from simple panel methods for rapid aerodynamic design~\cite{bock_2021_panel}, through traditional CFD for high-fidelity steady-state results~\cite{volpiani_2024_coda_aircraft_simulations}, to time-resolved simulations, or CSM added for aeroelasticity simulations~\cite{reimer_2015_flowsim} and multi-disciplinary optimizations~\cite{goertz_2017_mdo}. Realizing such gains in practice requires simulations across multiple fidelity levels, from rapid aerodynamic assessment to high-fidelity aeroelastic and multidisciplinary analyses. This creates further challenges regarding computational cost, uncertainty quantification, and the integration of life-cycle-oriented sustainability metrics into design optimization. Continued advances in CSE are therefore essential not only to design more efficient aircraft but also to evaluate such designs in a robust and scalable way.

\subsection{Designing Next-Generation Wind-Energy Parks}

Offshore and onshore wind energy can contribute substantially to meeting the growing global demand for clean energy, but realizing this potential requires continued expansion, cost reduction, and scientific innovation~\cite{veers2019grand_challenges}. CSE methods are used to improve the design and optimize operations. At the same time, we need to minimize local impact on the environment, e.g.,\ by reducing noise and vibration. This makes wind-energy systems another important example of how CSE can enable sustainability research in practice. Addressing these goals requires the integration of physical, environmental, and economic considerations on multiple scales. The questions of wind resource assessment, wake interaction, structural response, life-cycle assessment, and operation are tightly coupled and cannot be reliably addressed without advanced modeling, simulation, data assimilation, and optimization~\cite{veers2019grand_challenges}.

%More recently, surrogate strategies have been developed to make  life-cycle assessment and optimization of operations in wind parks computationally tractable \cite{lakdawala2024enhancing}.

By combining high-performance computing, data assimilation, and advanced numerical methods, CSE empowers accurate modeling of wind flows, with and without wake effects, under real-world conditions such as varying environmental forcing, geological constraints, and structural interactions. In recent years, the wind energy industry has increasingly embraced high-resolution meso- and microscale simulations for wind resource assessment and wake modeling, moving beyond traditional engineering models \cite{brown2023wake, fleming2017wake_steering_fieldtest}. 
Modern wind energy simulation relies on a hierarchy of computational tools, ranging from high-fidelity CFD and large-eddy simulation frameworks such as OpenFOAM~\cite{weller1998tensorial}, PALM \cite{palm-2020}, and ExaWind \cite{min2022exawind}, to reduced-order wake and optimization models such as FOXES \cite{Schmidt2023}. Related work compares simulation approaches of different fidelity for wind-turbine load prediction and evaluates a computationally inexpensive flow-based load indicator against high-resolution large-eddy simulations~\cite{bock2026simulation_methods}.
% and Whiffle (comment Julia: no reference found)
Together, these tools enable prediction, optimization, and control of turbine wake interactions across a wide range of spatial and temporal scales. This makes it possible to account turbine effects, complex terrain, and atmospheric stratification across various scales. As a result, CSE has contributed to more efficient site assessment, improved layout optimization, and better prediction of wake-induced power losses, thereby supporting improved wind-farm layouts, as illustrated by~\cite{mosetti1994windfarm_layout}.

Nevertheless, existing workflows often face trade-offs between computational speed and model fidelity. While these CFD-based methods provide higher accuracy, they remain computationally intensive. As a result, they do not scale easily for real-time prediction, large wind-farm configurations, or long-term scenario analysis at the desired scale. Many current workflows also lack integrated uncertainty quantification and real-time capability, both of which are crucial for robust decision-making in uncertain marine and complex-terrain conditions. To effectively support industry and policy makers, ongoing research and development in CSE must prioritize interoperable platforms, hybrid modeling approaches that combine physics-based and data-driven methods, and scalable digital twin infrastructure capable of adapting to evolving operational and meteorological regimes. These advances are essential to fully harness the transformative potential of CSE in real-world decision-making and innovation.

CSE provides the computational foundation to meet the challenges of expanding wind energy in a way that is uncertainty-aware, life-cycle-aware, and operationally robust. With sustained investment in methods, models, and computing infrastructure, CSE will remain central to shaping an efficient and sustainable offshore and onshore wind future.

\section{Toward Sustainability in CSE} 
\label{sec:toward_sustainable_cse}
\begin{figure}[t]
   \hspace{-3mm}\includegraphics[width=15cm]{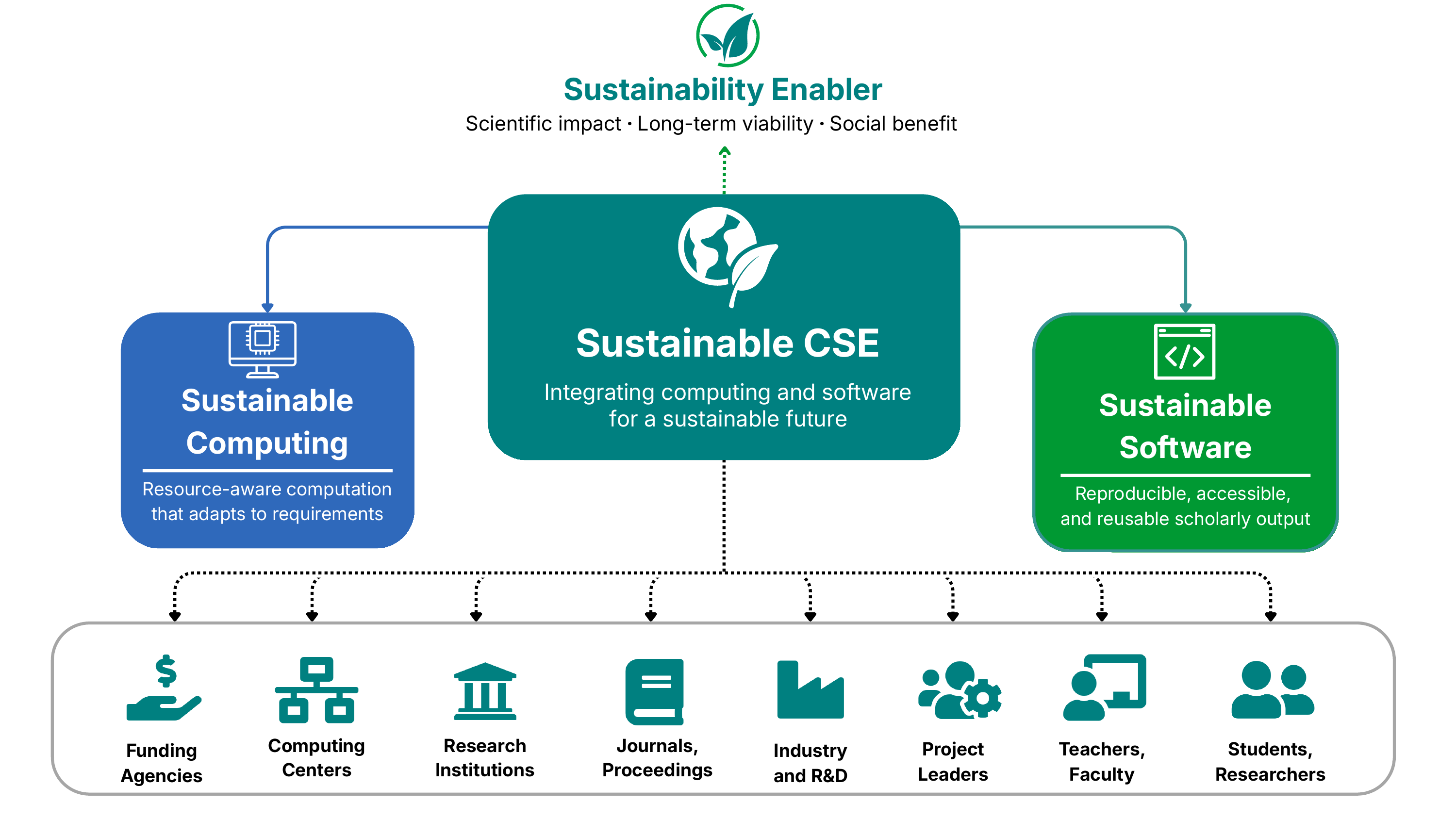}\hphantom{m}
    \caption{To enable simulation for a sustainable society, two components must become central within computational science and engineering: (i) \emph{sustainable computing} guarantees that computing itself, from numerical methods to hardware usage, is resource-aware and can adapt optimally to requirements; and (ii) \emph{sustainable software} guarantees that the main scholarly output of CSE remains reproducible, accessible, and reusable within the community and beyond. Together, these two aspects make CSE more sustainable as a field and secure its relevance for future scientific and technological endeavors. This development requires support from, and interaction among, stakeholders ranging from funding agencies to the researchers pushing the frontiers of CSE.}
    \label{fig:SCSE}
\end{figure}

The design of next-generation aircraft and next-generation wind farms are just two examples, among many across scientific disciplines, in which CSE methods and tools are used to optimize products or improve operations for a sustainable society. However, to act as an enabling technology for sustainable product development and decision support, CSE must also reflect on sustainability within its own practices. CSE cannot credibly enable sustainability in other domains if its own computational practices, software ecosystems, and knowledge flows are themselves unsustainable. 
%However, in recent years, CSE has faced challenges as a discipline to fully live up to its potential and expectations. 
We identify the following three aspects of sustainability within CSE:

%In the broader context, sustainability is typically regarded in the context of decision making regarding Environmental, Social, and Governance objectives, with plenty of potential to apply CSE as a tool. Although CSE is hence an important lever in sustainability research, sustainability is not addressed as a dimension within CSE itself. The planning, design, deployment, execution and operation of numerical simulations in CSE, however, constitute resources themselves that have to be considered from a sustainability point of view. There would be a clear benefit to systematically considering sustainability in the context of CSE because

\begin{itemize}
    \item \textbf{Energy footprint of CSE:} The application of CSE methods and algorithms has an energy footprint that grows with increasing data volumes, computational complexity, and problem sizes;
    \item \textbf{Longevity and reuse of CSE knowledge:} Fragmented access to methods, software, data, and digital infrastructure hinders the reuse, transfer, and efficient further development of existing advances;
    %\item CSE as a field has to assert itself against physics-agnostic artificial intelligence (AI) and machine learning (CSE has to sustain as a field).
    \item \textbf{Future relevance of CSE as a field:} Encoding physical principles, mathematical structure, and error management remains essential for applications and users but has become less visible in the context of recent data-driven and physics-agnostic approaches.
\end{itemize}

%Together, these aspects constitute sustainability dimensions within the CSE itself.
These three concerns do not translate into three parallel pillars. The first two define the operational pillars of sustainable computing and sustainable software, while the future relevance of CSE is an outcome they jointly support. Sustainable computing addresses the energy footprint of CSE, whereas sustainable software supports the dissemination, reuse, and preservation of CSE knowledge. Together, these pillars constitute what we refer to as \emph{sustainable CSE}, as illustrated in Figure~\ref{fig:SCSE}.

\section{Sustainable Computing}
\label{sec:sustainable_computing}  

The energy- and resource-efficiency aspects of sustainable computing are often discussed under the term green computing. The goals are to reduce environmental impact and maximize energy efficiency during any type of computing \cite{pereira2017energy_languages}. Ultimately, the goal is to find the best trade-off between energy consumption and algorithmic results. Energy-efficient computing matters for all classes of computational systems, from large-scale data centers to edge and handheld devices, because even small savings can have substantial aggregate effects. Avoiding redundant computation often reduces both runtime and energy consumption. Runtime is therefore a useful first approximation when average power remains comparable, since energy equals average power times runtime. GPU acceleration or vectorization can change power demand, however, so shorter runtime does not necessarily imply lower energy consumption.

Within CSE, we can view sustainable computing as optimization across three interconnected levels: hardware efficiency, algorithmic efficiency, and model efficiency. While hardware minimizes the cost of elementary operations, algorithms minimize the computational effort for a given task, and models minimize the computational effort required to generate scientifically relevant insight. Hardware efficiency depends not only on using a given computing system efficiently, but also on selecting an architecture suited to the workload. Vectorizable workloads may benefit from GPU acceleration, while specialized tasks may be better suited to architectures such as field-programmable gate arrays or neuromorphic systems. Exploiting these alternatives requires software that can target different architectures without forcing users to reformulate the underlying scientific model. Domain-specific languages can provide this flexibility. For example, NEST and its modeling language NESTML allow model components to be specified at a high level and executable code to be generated for deployment on CPUs, GPUs, and neuromorphic platforms such as SpiNNaker~\cite{gewaltig2007nest,linssen2025nestml,furber2014spinnaker,mayr2019spinnaker2}.

At the algorithmic level, mixed-precision methods can reduce computational and memory costs by performing selected operations at lower precision while retaining higher precision where required for numerical reliability~\cite{higham2022mixed_precision}. Compiler-based numerical rewriting can help explore precision choices and alternative expressions while assessing their effects on accuracy. Related tools include Poseidon for automated floating-point optimization~\cite{qian2026poseidon} and RAPTOR for numerical profiling and precision exploration~\cite{hoerold2025raptor}. At the infrastructure level, however, sustainability cannot be assessed from job-level energy consumption alone. HPC systems also entail embodied environmental impacts through their manufacture, installation, upgrading, and end-of-life treatment, while their total cost of ownership includes acquisition, energy and cooling, maintenance, staffing, and decommissioning. System architecture, service life, and useful utilization must therefore be considered alongside the operational efficiency of individual computations. In particular, concentrating scientifically valuable workloads on fewer, well-utilized systems may reduce the cost and environmental impact per useful computation compared with operating additional underutilized infrastructure.

Finally, transparency about computational resource consumption is essential, from individual algorithms to complete computational workflows. In the following, we discuss key levers for managing resource consumption and conclude the section by summarizing approaches for monitoring and assessing computational efficiency.

%Activists are actively researching the precision with which numbers are represented \cite{higham2022mixed_precision} and the precision with which algorithmic results are needed  \cite{vassiliadis2016automatic_significance,nikolopoulos2014significance_based_computing}.

%Algorithms can be selected according to their resource consumption, such as switching from a less performant, linear, search algorithm to a fast, hashed or indexed, search algorithm. Within high-performance computing, the Green500-list is a biannual ranking of supercomputers, from the top-500 list of supercomputers, in terms of energy efficiency measuring performance per Watt. Resource demand has gained renewed interest in the context of machine learning and artificial intelligence, where the training of deep artificial neural networks and LLMs, including generative AI that consumes large amounts of energy resources  \cite{de2023growing,fernandez-etal-2025-energy}. This also raises questions about the sustainable usage of machine learning in CSE.

%Finally, transparency about computational resource consumption is essential, from individual algorithms to complete computational workflows. In the following, we discuss key levers for managing resource consumption and conclude the section by summarizing approaches for monitoring and assessing computational efficiency.

%, and second, by using CSE methods themselves as levers to reduce this consumption in a goal-oriented way. 

\subsection{Monitoring of Resource Consumption in High-Performance and Desktop Computation}

Monitoring and predicting the energy demand of computations both in high-performance computing (HPC) and desktop computation is an important prerequisite for resource-aware CSE workflows \cite{Czarnul2019,Ge2012,Ezzatti2019PowerAware}. 
% AC
Meaningful sustainability assessment benefits from measurement at multiple scales. Within a single running job, tools such as LIKWID access hardware performance counters to attribute energy use to individual code regions~\cite{Treibig2010Likwid}. At the level of a full application or workflow, Alumet standardizes the measurement of energy consumption across heterogeneous components~\cite{raffin2025Alumet}. At the infrastructure level, systems such as CEEMS and EAR monitor and manage energy consumption across entire clusters or data centers~\cite{Paipuri2024CEEMS,Corbalan2025EAR}.

Where direct measurement is not feasible or too costly, prediction-based approaches complement these tools. One recent example is the framework \emph{UoPC: A User-Based Online Framework to Predict Job Power Consumption in HPC Systems}. UoPC uses a non-parametric model based on individual users' job histories and therefore requires neither model training nor a large system-wide training data set. On data from more than 1.3 million jobs executed on the Japanese supercomputer Fugaku, it achieved a mean absolute percentage error of 10\% for both average and maximum job power consumption, with an average inference time of approximately 0.08 seconds per job~\cite{antici_uopc_energy_monitoring}.

% One recent example is the framework \emph{UoPC: A User-Based Online Framework to Predict Job Power Consumption in HPC Systems}~\cite{antici_uopc_energy_monitoring}. 

% UoPC predicts job power consumption in HPC systems by leveraging ML-based models tailored to individual users, thereby reducing the need for voluminous training data. This makes the framework practical for diverse high-performance computing environments and workloads. It was evaluated on the Japanese supercomputer Fugaku using more than 1.3 million jobs and achieved a prediction error of only 10\%, with minimal overhead for system operations~\cite{antici_uopc_energy_monitoring}.

Frameworks of this kind make it possible to integrate energy awareness more systematically into everyday computational practice. For example, they can be used to relate the cost of compute time more directly to job power consumption, thereby creating incentives for users to optimize their workflows not only with respect to runtime, but also with respect to energy demand. 
% More generally, meaningful sustainability assessment benefits from measurement and prediction at multiple levels, including infrastructure, applications, and individual algorithmic components.
In this sense, energy monitoring provides the transparency needed to make sustainable computing actionable in CSE.
%AC
% end Achim

\subsection{Resource-Aware Computing Through Adaptivity and Parallel Scalability}
\label{subsubsec:amr}

Adaptive discretization methods are a key building block of sustainable computing in CSE because they allocate computational effort only where needed. Adaptive mesh refinement (AMR) dynamically refines and coarsens the computational mesh, for example in the classical Berger--Oliger framework~\cite{berger1984amrhyperbolic}, while adaptive finite element methods use a posteriori error estimates to control the distribution of degrees of freedom~\cite{babuska1978aposteriori,verfurth1994aposteriori}. In both cases, adaptivity provides a systematic mechanism for balancing computational cost and solution accuracy.

A resource-aware perspective becomes particularly important when the objective is the accurate prediction of selected quantities of interest rather than uniform accuracy throughout the domain. Goal-oriented adaptivity combines local error estimates with sensitivities of the target functional to concentrate computational effort where it most affects the simulation objective~\cite{becker2001optimalcontrol}. Moreover, convergence theory establishes adaptive refinement as a mathematically rigorous strategy rather than a heuristic procedure~\cite{morin2002convergence}. Together, these methods systematically balance accuracy and computational cost.

Realizing these benefits, however, requires efficient implementation. Adaptive methods rely on scalable data structures, load balancing, and solver strategies to prevent refinement overhead from outweighing computational savings. Domain decomposition methods, including overlapping Schwarz techniques and adaptive coarse-space approaches, provide complementary mechanisms for distributing nonuniform workloads efficiently on large-scale systems~\cite{dryja1994domaindecomposition,heinlein2020adaptivegdsw}. 
Fine-grained partitioning and dynamic load balancing on hierarchical meshes provide one example of how parallel performance can be improved in locally adaptive coupled simulations~\cite{wegmann2025parallel_coupling}.
In practice, sustainable adaptivity therefore combines numerical decisions, such as refinement criteria and solver tolerances, with efficient parallel execution.

By avoiding unnecessary computation in regions of limited relevance, adaptive methods can substantially reduce time-to-solution, energy consumption, and computational footprint while maintaining the desired accuracy. Achieving these benefits in everyday CSE practice requires reusable software infrastructures that expose adaptive capabilities through flexible, portable, and scalable interfaces. Such enabling frameworks lower the barrier to adopting resource-aware methods and thereby propagate sustainable computing practices across a broad range of applications.

\subsection{Resource-Aware Computing Through Algorithmic Differentiation}\label{subsubsec:ad}

Algorithmic differentiation (AD) is a particularly relevant example of resource-optimized
computing in CSE because it enables sensitivities and gradients to be computed accurately and
systematically without the need for large amounts of hand-derived derivative code or the use
of finite differencing \cite{GriewankWalther2008, Naumann2012}. In settings such as
multidisciplinary analysis and optimization, this can reduce development effort, avoid
duplicated implementations, and make advanced optimization workflows more reliable and
scalable. In that sense, AD contributes to sustainability not only through computational
efficiency, but also through improved maintainability and extensibility of numerical software.

The resource efficiency of AD becomes particularly important through its two complementary
modes of operation. Forward (tangent-linear) mode propagates sensitivities alongside the primal
computation and is efficient when derivatives with respect to few inputs are required, whereas
reverse (adjoint) mode computes the gradient of a scalar objective with respect to very many
parameters at a cost that is essentially independent of their number. It is this adjoint
capability that turns the accurate evaluation of high-dimensional gradients from a prohibitive
expense into a routine operation. Beyond these basic modes, techniques such as checkpointing,
exploitation of sparsity, and higher-order accumulation extend AD to large and long-running
simulations, and the corresponding methods and tools continue to advance as an active field of
research \cite{Forth2012RecentAdvances}.

This capability is what makes several resource-intensive CSE workflows tractable in the first
place. In gradient-based design optimization, a single adjoint evaluation provides the
sensitivities of an objective functional with respect to all design variables, so that the cost
of one optimization step becomes largely decoupled from the dimension of the design space
\cite{GilesPierce2000}. In uncertainty quantification, the same derivative information can be
propagated to approximate moments and sensitivities of quantities of interest, offering an
efficient alternative to purely sampling-based approaches when the number of uncertain
parameters is large \cite{Maldonado2019ADUPROP}. In both cases, AD reduces the number of
expensive model evaluations required and thereby lowers the overall computational and energetic
cost of the workflow rather than of a single run.

Realizing these benefits sustainably, however, requires AD to be available as reusable and
maintainable software infrastructure rather than as a one-off implementation effort. Modern
compiler-based tools such as Enzyme perform differentiation directly on optimized intermediate
representations and can therefore provide efficient derivatives across many programming
languages without rewriting existing code \cite{MosesChuravy2020Enzyme}. Such approaches keep
differentiability a durable property of a code base as it evolves, and they connect naturally to
the software examples discussed below. In this way, AD illustrates how resource-optimized
methods and sustainable software reinforce one another in the pursuit of sustainable computing.

\subsection{Model Hierarchies as a Lever for Sustainable Computing}
\label{subsubsec:model_hierarchies}

Many CSE application domains naturally admit model hierarchies, ranging from inexpensive, low-fidelity descriptions to expensive, high-fidelity simulations. Exploiting these hierarchies in a systematic way is a powerful lever for sustainable computing, because it allows researchers to reserve the most energy-intensive models for situations where they provide the highest information gain. Typical examples include using simplified physics, reduced spatial or temporal resolution, lower-order numerical schemes, or reduced-order and surrogate models in early design and screening phases \cite{kowalski2017,scholz2024dispersion,benner2022model_reduction,kichler2024sobolev_pruning}, while deploying high-fidelity simulations only for final verification, calibration, or the analysis of critical scenarios. Model hierarchies can be exploited to control the fidelity level at which processes are resolved in a spatially distributed manner. For instance, a computational region of primary interest to a hazard impact simulation or a flow field simulation can be resolved at high fidelity, while the remainder of the domain is represented by a less expensive reduced model, as illustrated in~\cite{maso2022coupled_models,lovholt2008tsunami}.

Such multi-fidelity strategies can substantially reduce computational and energetic cost at the level of complete workflows rather than single runs. However, they require careful coupling between models, consistent error and uncertainty assessment across fidelity levels, and software infrastructures that make it easy to switch between or combine models within one environment. Embedding model hierarchies into CSE practice therefore contributes to sustainable computing not only by lowering resource consumption, but also by structuring simulations in a way that makes the relationship between cost and insight more transparent and controllable. Model hierarchies thus provide another methodological lever for matching fidelity and resource consumption to the actual objective of the workflow.

\subsection{Key Takeaway}

Taken together, these examples show that sustainable computing in CSE extends beyond measuring energy consumption or optimizing individual computations. It requires the coordinated design of models, algorithms, software, and computing infrastructure to balance scientific benefit, resource consumption, and time-to-solution across entire computational workflows. Further research should develop and assess resource-aware strategies such as mixed-precision computation, low-rank and tensor methods, sparsity exploitation, and communication-avoiding algorithms. Sharing reusable simulation data can complement these efforts by reducing unnecessary recomputation. The Johns Hopkins Turbulence Database provides a prominent example of this approach~\cite{li2008public_turbulence_database}.

\section{Sustainable Software}
\label{sec:sustainable_software}   

Software is at the heart of CSE, turning ideas into practice. With the emergence of large language models and agentic systems we observe a rapid change of how software is produced. Agentic systems can already inspect repositories, modify multiple files, execute commands and tests, and interact with external resources as part of extended development workflows. These capabilities may accelerate research software development, but they do not reduce the need for maintainability, verification, documentation, and human responsibility. On the contrary, the future impact of the field will depend strongly on the sustainable development, integration, creative use, verification, and certification of complex research software. 

Software represents a substantial investment: it enables reproducible scientific discovery, integrates outcomes across disciplines, supports collaboration, and provides a foundation on which new methods and applications can build. Consequently, reliability, reusability, composability, maintainability, and verifiability are becoming increasingly important across applied science and engineering. One example of the long-term impact of research software is given by the work of LeVeque and collaborators, who have developed a family of freely available numerical software over several decades and advocated for reusing software and reproducibility ever since  \cite{leveque2012reproducible,leveque2013top}. Clawpack \cite{mandli2016clawpack} provides robust implementations of finite-volume methods for wave propagation, while extensions such as GeoClaw \cite{berger2011geoclaw}, AMRClaw \cite{berger1998adaptive}, and ForestClaw \cite{Calhoun2017ForestClaw} have broadened this foundation toward geophysical applications, adaptivity and parallel scalability. This ecosystem illustrates how sustained software development can make methodological advances accessible, reusable, and extensible across research, education, and applications.

In the following, we present four software projects that highlight different facets of sustainable software in CSE, including maintainability, openness, testing, portability, extensibility, performance, and community uptake.
The four examples were selected as illustrative, actively developed projects represented within the author group. Notably, they are not intended as a comprehensive survey or ranking of sustainable CSE software. We organize the examples into three categories, which are not mutually exclusive and serve as different perspectives on sustainable CSE software.

\subsection{Method-Development Software}

Method-development software provides environments in which numerical methods can be implemented, tested, compared, and transferred into research practice.\\

\noindent \emph{Example 1: Hyperbolic Conservation Laws: Trixi.jl}: Trixi.jl \cite{ranocha2022adaptive,schlottkelakemper2021purely,schlottkelakemper2020trixi} is a numerical simulation framework that provides adaptive high-order methods for conservation laws and therefore serves the scientific application layer. The project was started in 2020 by a small team of four developers including two postdocs and two professors, and has since evolved into the Trixi Framework, offering several related open-source software packages on GitHub~\cite{Trixi}.
%The main repository has gathered 700 GitHub stars, and 
By now, Trixi.jl has been adopted in publications beyond the core development team, indicating use as a community resource. The origin of Trixi.jl as research software in CSE is somewhat unusual. Its primary motivation was not research alone, but to develop a tool that is useful for both research and teaching numerical methods. All principal developers had already been engaged in high-performance simulation software, yet these highly optimized codes, written in traditional programming languages such as C, C++, and Fortran, were often ineffective for teaching purposes. A master’s student typically spent substantial time setting up the software and its dependencies and understanding the code, leaving little time for research or contributions. Thus, the three primary design goals of Trixi.jl were established from the outset to be extensibility for research and teaching, ease of use for students and collaborators, and performance, with extensibility leading in emphasis. These goals guided decisions such as using the Julia programming language \cite{bezanson2017julia} and sometimes prioritizing code readability over ultimate abstraction, e.g.,\ accepting separate code paths for 1D, 2D, and 3D problems. This allows new users and collaborators to focus on their specific tasks without having to worry about the general case from the outset. At the same time, Trixi.jl’s performance is at least comparable to mature HPC code written in traditional programming languages, and it has been shown to scale well to more than \num{100} GPUs or \num{10000} CPU cores on modern high-performance systems \cite{ranocha2023efficient}. This computational efficiency contributes to sustainability by reducing the time, energy, and computational resources required to obtain scientific results at scale. Several software-design decisions that are not necessarily typical of research software, but appear essential from the current perspective, include an open-source approach from the outset, extensive integration tests covering over \SI{97}{\percent} of the code, a comprehensive collection of examples, and ample documentation for new users and developers. Thus, Trixi.jl has implemented the FAIR/FAIR4RS principles \cite{wilkinson2016fair,barker2022introducing,deveaughgeiss2025climate_fair_foss} from the beginning. Using the Julia language has also simplified getting started with Trixi.jl \cite{ranocha2022adaptive} and integrating other open-source packages. Its native C interoperability and ability to distribute precompiled software libraries allow users to utilize the C/C++ packages \texttt{p4est} \cite{burstedde2011p4est} and \texttt{t8code} \cite{holke2025t8code} for parallel adaptive mesh management, thereby saving development time. Although relying on third-party software can lead to some overhead as upstream packages evolve, the ability to focus on the project's core goals has ultimately been a net win for the Trixi.jl community. Equally important, continuous exchange with the broader Julia and scientific software communities has shaped many aspects of Trixi.jl, making it possible to build on and contribute back to a vibrant ecosystem rather than developing the framework in isolation.\\

\noindent\emph{Further examples of sustainable method-focused software:} Many other successful software projects exist that have a strong focus on supporting the development of various numerical methods. DUNE provides durable abstractions and modular interfaces for numerical method development~\cite{bastian2021dune,bastian2008dune_grid}, while deal.II provides a best-practice example through reusable finite-element abstractions, extensive documentation and testing, and long-term community development~\cite{dealii2019design}. Clawpack similarly illustrates how a long-lived open ecosystem can connect method development, documentation, research, and education~\cite{mandli2016clawpack,leveque2002finite_volume}. 

\subsection{Application-Focused Software}

Application-focused software sustains domain-specific modeling and analysis capabilities while integrating methods, infrastructure, and community knowledge.\\

\noindent \emph{Example 2: Computational fluid dynamics with CODA}: Computational fluid dynamics (CFD) codes for complex industrial applications such as aircraft design and certification are facing a growing number of tasks and challenges. Addressing these challenges while providing a common, sustainable, and extensible software platform for next-generation industrial CFD is the primary objective of the CODA code. CODA is the CFD software being developed as part of a collaboration between the French Aerospace Lab ONERA, the German Aerospace Center (DLR), Airbus, and their European research partners~\cite{volpiani_2024_coda_aircraft_simulations}. The CODA collaboration addresses four related challenges:

\begin{itemize}
    \item Sustainable adaptation to increasingly complex and fragmented computing hardware, with accelerators and deep cache hierarchies being the norm of today's systems;
    \item Efficient methods addressing the growing gap between memory bandwidth and available performance, also referred to as the memory gap, which favors high-order methods with high arithmetic intensity over low-order methods with low arithmetic intensity;
    \item Effective access to simulation sensitivities and scalable in-memory transfer of solution data for multidisciplinary design, analysis, and optimization (MDAO) on HPC systems;
    \item Modular and well-structured rather than monolithic software to enable verifiability and testability, both key drivers of sustainable software development.
\end{itemize}

Some of these capabilities can be retrofitted into a legacy code base, e.g.,\ TAU~\cite{gerhold_1997_tau}, while others, such as correct sensitivities, are harder to achieve. The Flucs~\cite{leicht_2016_flucs} code started at DLR with modular code design as its main concept and has since become \emph{CFD by ONERA, DLR and Airbus}~(CODA)~\cite{huismann_2021_hypercoda}, co-developed and jointly owned by the three partners. Its main design goal was not a stand-alone CFD solver, but a library that can exchange simulation results in-memory with other codes. CODA is therefore integrated into the Python-based, HPC-capable FlowSimulator framework~\cite{reimer_2015_flowsim,huismann_2021_flowsim}, which allows the exchange of arbitrary data on and between grids for MDAO processes, including the design of sustainable aircraft as discussed in Subsection~\ref{subsec:aircraft}. The sensitivities required for gradient-based optimization are computed using algorithmic differentiation, see Subsection~\ref{subsubsec:ad}, allowing any combination of discretization, partial differential equations, and convection scheme. Because CODA focuses on steady-state Reynolds-averaged Navier–Stokes equations (RANS) calculations for transport aircraft, its strong non-linear solvers place a high load on sparse linear algebra; CODA therefore uses the linear solver library Spliss~\cite{krzikalla_2021_spliss}, whose state-of-the-art techniques for sparse matrices can leverage accelerator cards because, compared with highly non-linear residual evaluation and system-matrix calculation, sparse linear solvers exhibit regular memory patterns and predictable code paths~\cite{wagner_2024_spliss_gpu}.\\

\noindent \emph{Further examples of sustainable application-focused software:} Other application-focused projects with a strong CFD focus include FLEXI and GAL\AE XI, which combine high-order simulation capabilities with performance-portable implementations for modern accelerator systems~\cite{krais2021flexi,flexi_software_v25_10_0,kurz2025galaexi,starr2026architecture_agnostic,galaexi_software_1f574224}. 
%m-AIA is a multi-physics PDE solver framework for efficient HPC simulation of coupled CFD, aeroacoustic, and multiphase-flow problems using multiple numerical schemes on hierarchical Cartesian meshes \cite{maia_software_2026}. 
Bryne illustrates how application-focused software can build directly on method-development infrastructure: using the DUNE-FEM Python API, it turns finite-element prototypes into reusable, metadata-enriched simulations and supports operator-split multiphysics coupling~\cite{terschanski2026bryne,bryne_software_v1_1_0}. The software tool UG was crucial for several scientific discoveries, in particular in the biosciences \cite{vogel2013ug4}.
Beyond CFD, CADET-Core provides maintained and versioned process-modeling capabilities for biotechnology~\cite{leweke2025cadet_core,cadet_core_software_v5_1_1}, while NEST and NESTML combine a neuroscience simulation environment with a domain-specific modeling language~\cite{gewaltig2007nest,linssen2025nestml}. In energy-system analysis and sustainability assessment, SecMOD-MILP, \texttt{bw\_timex}, \texttt{optimex}, and \texttt{pbaesa} support reusable workflows for optimization and life-cycle assessment~\cite{reinert2023secmod_milp,secmod_milp_software_v2_0_0,muller2025time_explicit_lca,bw_timex_software_v1_4_0,diepers2026time_explicit_optimization,optimex_software_v0_7_0,hartmann2026allocation,pbaesa_software_v0_1_2}.

\subsection{Reusable Frameworks and Enabling Infrastructure}
%% from the website
Reusable frameworks and enabling infrastructure support broad classes of models and coupled applications through shared capabilities such as multiphysics composition, mesh management, adaptivity, partitioning, coupling, and performance portability. As computational tasks grow, these capabilities enable applications to exploit large-scale and emerging HPC systems without redeveloping common infrastructure.\\

\noindent \emph{Example 3: Block-Structured Adaptivity: AMReX}: AMReX is a publicly available software framework~\cite{AMReX} designed to build massively parallel block-structured AMR applications that implement multi-physics simulations including those sketched in Section~\ref{subsubsec:amr}. AMReX originated in the 1990s as BoxLib. Under funding from the U.S. Department of Energy's Exascale Computing Project it became AMReX, a general-purpose future-looking framework for simulation codes wanting to use block-structured mesh refinement on machines from laptops to supercomputers without having to support multiple versions of their codes. AMReX provides a performance portability layer and a number of supporting constructs to shorten the time to solution of existing applications as well as time to development of nascent applications. It provides distributed data containers, efficient iterators over the data, and GPU-optimized parallel communication, domain decomposition, load balancing and memory management capabilities.
AMReX also illustrates sustainable code development through early architectural decisions that avoid unnecessary technical debt, learning from experience and abandoning approaches that no longer serve the project, and designing beyond the interests of the current developer and user base. Such strategic evolution helps software adapt to changing hardware and software stacks while retaining backward compatibility. It also helps projects balance generality with specificity without becoming diffuse, and treat performance and portability as complementary rather than competing goals. These properties are central to the long-term value of software infrastructures in science in general, but are of particular importance in CSE.\\

\noindent \emph{Example 4: Versatile Mesh Adaptation: t8code}: Whereas AMReX integrates adaptive mesh management into a broad framework for scalable simulations, the mesh adaptation library \texttt{t8code}~\cite{holke2025t8code} provides focused, reusable mesh adaptation and partitioning capabilities that offer applications a direct route into scalable HPC environments. Developed at DLR, \texttt{t8code} features a unique, scalable tree-based refinement strategy and a scalable and extremely fast partitioning approach based on space-filling curves. It serves as a mesh management backend for scientific and industrial applications in manufacturing, visualization, energy, climate, and flood simulations. Its development infrastructure combines elaborate change and release management with software testing, architecture documentation, and automated continuous benchmarking of scalability and efficiency~\cite{KoLaKi-2025-SystemsBenchmarking}.\\

\noindent\emph{Further examples of sustainable and reusable enabling software frameworks:} Related enabling software separates reusable mesh capabilities from application-specific solvers. Within the Clawpack ecosystem, AMRClaw packages adaptive-refinement algorithms for hyperbolic problems, while ForestClaw combines patch-based methods with scalable forest-of-quadtrees mesh management~\cite{berger1998adaptive,Calhoun2017ForestClaw}. Complementary frameworks emphasize modularity and coupling as enabling infrastructure. DUNE supplies reusable numerical interfaces across application domains~\cite{bastian2021dune,bastian2008dune_grid}. The coupling framework preCICE enables partitioned multiphysics coupling between independently developed simulation codes~\cite{chourdakis2022precice}. Micro Manager supports adaptive, flexible two-scale coupling~\cite{desai2023micro_manager}. Portability infrastructure can also reduce repeated porting effort: Kokkos provides a programming model for heterogeneous architectures~\cite{trott2022kokkos}, LLVM OpenMP offload optimizations support its GPU execution~\cite{gayatri2024kokkos_offload}, and Polygeist enables GPU-to-CPU translation~\cite{moses2023polygeist}. Together, they show how shared interfaces can reduce duplicated implementation effort while allowing specialized solvers to evolve independently.

\subsection{Key Takeaway} 
Taken together, these actively developed projects show that sustainable software in CSE cannot be reduced to a single class of tools or a fixed position in the software stack. Method-development software must make new numerical approaches testable, extensible, and transferable; application-focused software must remain reliable, verifiable, and adaptable to evolving scientific and hardware requirements; and reusable frameworks and enabling infrastructure must provide stable interfaces, portability, and scalable capabilities that avoid repeatedly rebuilding common foundations. The four illustrative examples also demonstrate that these roles overlap: applications depend on shared infrastructure, while frameworks acquire lasting value through adoption, extension, and feedback from scientific communities. Sustainable CSE software therefore emerges from the joint design and long-term stewardship of methods, applications, and infrastructure, supported by testing, documentation, governance, and research software engineering.

In this context, research software engineering (RSE) provides an important response to the growing role of software in computational research~\cite{Felderer25, code4science}. While RSE is a broader development shaped by many software-intensive research domains, it is highly relevant to CSE because it helps transfer established software engineering practices into research environments~\cite{fritsch2026gi_derse_guideline}, supports training and education tailored to CSE communities~\cite{Chourdakis_Ashraf_Narvaez_Rivas_Neckel_Bungartz_2025,Bertrand2025RSETeaching2025}, and connects bottom-up community building with top-down software policies and institutional support~\cite{report_RSE_policy}. At the institutional level, central RSE units offer one organizational model for putting such support into practice by combining sustained software infrastructure with mentoring and support for research projects~\cite{kempf2026central_rse_units}.
When combined with strong domain-specific and methodological expertise, RSE provides the organizational framework through which sustainable software practices can be established and maintained.

As software is increasingly assembled, modified, and reused through both manual and agentic workflows, provenance and attribution also become sustainability concerns. The software components and intellectual contributions on which generated or modified code depends must remain visible, identifiable, and citable~\cite{smith2016software_citation}. Agentic development therefore increases the need for persistent, version-specific software references and machine-actionable metadata rather than making established software-engineering and citation practices obsolete.

\section{Integrating Sustainability in CSE}
\label{sec:integration}

The preceding sections have considered sustainability in CSE from two complementary perspectives: CSE as an enabler of more sustainable products and engineering systems, and the sustainability of CSE itself, particularly with respect to computational resource use and the long-term availability of research software and CSE knowledge. These perspectives are interdependent and deeply entangled. CSE can create credible and lasting benefits in application domains only if its methods are implemented in robust and reusable software and its computational resources are used deliberately.

This also means that sustainability cannot be reduced to minimizing computational activity. For example, performing additional MDAO simulations increases computational energy consumption but may still yield a net environmental benefit if the resulting product or process achieves environmental savings over its life cycle that exceed the additional impact of the simulations. Similarly, developing maintainable and reusable research software may require greater effort initially but can reduce duplicated work, enable reproducibility, and preserve methods and knowledge over time. Sustainable CSE therefore calls for a long-term perspective that balances, across these different timescales, immediate resource consumption and development effort against anticipated information gain and scientific value, potential downstream environmental benefits, and the development of long-term capabilities. \textbf{At a time of rapidly growing demand for computation, this perspective presents an opportunity for CSE to advance methods, software, and practices that generate scientific insight using only the computational resources necessary to achieve the intended scientific and societal benefits.}

To date, however, a major challenge is that the immediate costs of sustainable practices are often borne by individual researchers and projects, whereas many of their benefits emerge only later or accrue to the wider scientific community and society. Individual action alone is therefore insufficient. Sustainable CSE requires coordinated incentives, infrastructure, standards, and support from the stakeholders who shape how computational research is funded, conducted, disseminated, and evaluated. The following section translates these considerations into stakeholder-specific recommendations.

\section{Recommendations}
\label{sec:recommendations}

As a group of CSE researchers, we propose a concerted effort within the CSE community to establish realistic and achievable metrics and best practices that embed sustainability as a core design principle in CSE activities rather than treating it as an afterthought or side effect. The following recommendations combine bottom-up practices with top-down institutional support and emerged from the first Conference on Sustainable Computational Science and Engineering, held in Steinfeld, North Rhine-Westphalia, Germany, in 2025~\cite{scse2025conference}.

\subsection{Recommendations for Funding Agencies}

\hphantom\\\vspace{-12pt}

\setlength{\intextsep}{0pt}
\begin{wrapfigure}{l}{20pt} 
\centering 
\vspace{-1pt}
{\fontsize{20pt}{30pt}\selectfont \textcolor{teal}{\faHandHoldingUsd }} 
\end{wrapfigure}
\noindent  Like all disciplines, computational science and engineering relies strongly on external funding, which may come from public agencies or private entities. Securing these resources is typically a highly competitive process that gives funding agencies significant leverage and influence over activities in scientific research. Agencies funding CSE should:

\begin{itemize}
    \item \textbf{Extend and tailor the scope of available calls: }
    There is a gap between funding options for projects that facilitate the development of novel software assets and funding lines that support the further development of established large infrastructure-level software packages. There are limited funding opportunities that target software as a research objective, e.g.,\ to ensure its longevity and to improve its energy footprint.
    % next point seen critical therefore commented out and substituted by more concrete suggestion
    %\item request standardized software management plan with proposals
    \item \textbf{Treat software as an infrastructure asset in all phases of the proposal: }
    Proposals, reviews, and project evaluations should explicitly acknowledge software as a long-lived research output. This includes realistic planning for maintenance, documentation, interoperability, and community uptake beyond the immediate duration of the project.
    \item \textbf{Require software and its status to be listed in final reports: }
    Currently, the software assets developed within funded projects are not reported systematically in a way that facilitates searchability and visibility. Nevertheless, many projects list their software results. Requiring software results, their status, and licensing information to be listed in a standardized way, following specifications such as REUSE~\cite{reuse_specification_3_3,reuse_tool_software_v6_2_0}, would incentivize reuse and further development, encourage contributions to existing projects, and improve interoperability. In the long run, this would make research funding more effective.
    \item \textbf{Request statements on expected and actual energy consumption in proposals and reports: }
    In the near future, such statements should remain lightweight and proportional to the size of the project, not least because standardized assessment tools are not yet widely available. Nevertheless, requiring applicants to engage with the topic can help normalize energy awareness and encourage them to justify why their chosen computational approach is appropriate from both a scientific and a resource perspective. However, in the long term, quantitative assessments should become standard practice.
\end{itemize}

% Remark Julia: We have to be careful as we don't want responsible people at the funding agencies to take up ideas of new reporting opportunities in an unreflective way. I rather suggest to couple this to an incentive system, such as 'additional modules can be applied for if you commit to also track the energy footprint of the project or so'. I'm not so sure as of now how to bring this in, so conserve my thought here in the comments.

\noindent \textbf{What is the benefit for the funding agencies?}\\
Transparency regarding the energy footprint of funded projects can be used as a lever to guide future directions. Accessible information on software results facilitates reuse and makes research spending more effective and impactful in the long run.

\subsection{Recommendations for Supercomputing Centers}
\hphantom\\\vspace{-12pt}

\setlength{\intextsep}{0pt}
\begin{wrapfigure}{l}{20pt} 
\centering 
\vspace{-2pt}
{\fontsize{20pt}{30pt}\selectfont \textcolor{teal}{\faNetworkWired }} 
\end{wrapfigure}
\noindent  High‑performance computing centers play a central role in computational science and engineering, providing large-scale computational infrastructure. Their operation involves substantial energy use, making sustainability a critical consideration. At the same time, compute time is allocated competitively, giving the associated allocation policies considerable influence on the field. Computing centers for CSE research should:

\begin{itemize}
    \item \textbf{Revise compute-access criteria for HPC proposals: }
    Currently, the relevance of the research topic and demonstrated scalability are among the criteria for granting compute time. These should be complemented by a statement justifying why the chosen HPC approach is appropriate from an energy perspective and by information on the sustainability of the software.
    %\item require computation time proposals to motivate their approach to solving the problem in question with respect to energy
    %\item require computation time proposals to disclose the sustainability of their software
    \item \textbf{Commit to sustainability goals: }
    Supercomputing centers should provide transparent guidelines on sustainable software and sustainable computing requirements. This includes providing tools to enable energy footprint estimation and monitoring for individual projects. Another measure would be to offer training and tooling support that helps potential users understand and apply these measures.
    %\item provide tools to enable energy footprint estimation
    %\item reject or penalise proposals with inadequate motivations with regard to energy or substandard software sustainability concepts. 
    \item \textbf{Provide feedback to compute time applicants:}
    Where appropriate, this feedback should recommend alternative computational approaches, such as the use of GPU-accelerated HPC resources when these are likely to be more energy-efficient than CPU-only systems, or methodological changes that reduce resource use while preserving the scientific objective.
    Such recommendations should consider not only job-level energy consumption but also the efficient utilization and broader life-cycle costs of the available computing infrastructure.
\end{itemize}

\noindent \textbf{What is the benefit for the supercomputing centers?}\\
These measures would improve the use of the energy invested in computation, thereby increasing the reach and impact of the supercomputing center and raising overall productivity while reducing the energy consumption of individual jobs through more efficient computational approaches.

\newpage
\subsection{Recommendations for Research Institutions}
\hphantom\\\vspace{-12pt}

\setlength{\intextsep}{0pt}
\begin{wrapfigure}{l}{16pt} 
\centering 
\vspace{-0pt}
{\fontsize{20pt}{30pt}\selectfont \textcolor{teal}{\faUniversity }} 
\end{wrapfigure}
\noindent  Research institutions such as universities and national laboratories are the backbone of computational science and engineering, hosting the researchers and students that drive CSE. The recruitment, training, collaboration, and allocation of resources shape both the pace and direction of CSE progress. By setting long‑term priorities, these institutions strongly influence the future direction of the field. To support sustainable CSE, research institutions should:

%JK
\begin{itemize}
    \item \textbf{Invest in software infrastructure:} Research software represents long-lived assets, yet it is rarely funded as such. Institutions should invest systematically in sustainable software infrastructure, including shared repositories and testing and archiving services. Such structural investment protects the substantial scientific value embedded in institutional codes and prevents the recurrent loss of effort when projects and their staff move on.
    \item \textbf{Be transparent about energy footprints:} The energy footprint of computational work is largely invisible to the researchers who generate it. Together with computing centers, institutions should make this information more accessible and support practical energy-footprint estimation through tooling, training, and local expertise. This helps to embed energy awareness into everyday research practice. 
    \item \textbf{Foster software visibility:} Sustainable software depends not only on funding but also on community and continuity. Institutions should foster the visibility of software projects. Connecting emerging codes to established ones spreads good engineering practices and reduces duplicated effort. 
    \item \textbf{Recognize software contributions:} As long as software is invisible in career evaluation, researchers have little incentive to invest in its quality and longevity. Institutions should recognize software contributions in hiring, promotion, and annual performance assessments, on par with traditional publications, clarifying that software is a core part of the scientific mission.
\end{itemize}

\noindent \textbf{What is the benefit for the research institutions?}\\ 
Institutions that support sustainable CSE strengthen the quality, visibility, and longevity of their research portfolio. They also improve the return on infrastructure investments by enabling more reusable software, better-trained researchers, and more transparent use of computational resources.

\subsection{Recommendations for Journals, Conferences, and Workshops}
\hphantom\\\vspace{-12pt}

\setlength{\intextsep}{0pt}
\begin{wrapfigure}{l}{16pt} 
\centering 
\vspace{-2pt}
{\fontsize{20pt}{30pt}\selectfont \textcolor{teal}{\faBook }} 
\end{wrapfigure}
\noindent  Scientific journals, conferences and workshops contribute immensely to the visibility and credibility of CSE research. Through editorial policies and peer‑review standards, they influence not only the findings that reach the community but also the way aspects of sustainable CSE are valued. As long as journal and conference papers remain the main currency of academia, their decisions have a significant influence. Editorial boards, program committees, and publishers should:  

%Jens, PB: 
\begin{itemize}
    \item \textbf{Provide transparent guidelines:} Establish transparent and enforceable guidelines on software quality, sustainability, and reproducibility. Making these expectations explicit before submission helps authors prepare accordingly and gives reviewers a consistent basis for assessment.
    \item \textbf{Require code disclosure:} Require manuscripts to disclose and cite the software used in a persistent and version-specific form, or to justify why this is not possible, e.g.,\ due to intellectual property or export-control restrictions. This requirement should also cover software dependencies selected, incorporated, or modified through AI-assisted or agentic workflows. Citable software strengthens reproducibility and gives due credit to the tools underlying the reported results.
    \item \textbf{Integrate code into review:} Include code and software artifacts in the review process, so that reproducibility and software quality are assessed alongside the scientific content. Details and requirements of a code review must be formulated in a reasonable way to ensure efficiency and effectiveness.
    \item \textbf{Request improvements:} Ask authors to improve submissions that do not meet reasonable standards of software sustainability, documentation, or reproducibility, rather than treating these as optional. Consistent follow-through gradually raises the baseline quality of computational research across the field.
    %%\item encourage the publication of well-documented negative computational results, unsuccessful algorithmic approaches, and software limitations whenever they provide valuable evidence to the community.
\end{itemize}
\noindent \textbf{What is the benefit for the journal?}\\
These measures improve reproducibility, strengthen community trust, and enhance the publication venue's reputation for rigorous and durable computational research and ultimately improve the quality of the published research. 

\subsection{Recommendations for Industry and Industrial R\&D}
\hphantom\\\vspace{-12pt}

\setlength{\intextsep}{0pt}
\begin{wrapfigure}{l}{20pt} 
\centering 
\vspace{-0pt}
{\fontsize{20pt}{30pt}\selectfont \textcolor{teal}{\faIndustry }} 
\end{wrapfigure}
\noindent Industrial stakeholders and industrial R\&D units are the main users and
collaborators of methods and concepts developed in academia. They shape how CSE
methods are translated into model-based design and operational decision support
under real-world constraints such as cost, time-to-market, certification, and
maintainability. Their policies and practices therefore strongly influence
whether sustainability principles become embedded in applied CSE workflows at
scale. They should:

\begin{itemize}
    \item \textbf{Consider sustainability criteria:} Apply life-cycle-oriented sustainability criteria in the development, selection, and deployment of CSE workflows and digital tools. Weighing environmental and long-term maintenance costs alongside performance and time-to-market leads to computational assets that remain valuable well beyond their initial deployment.
    \item \textbf{Support software ecosystems:} Support interoperable software ecosystems and collaboration models that enable effective knowledge transfer between academia, research centers, and industry. Building on shared, well-maintained components reduces duplicated effort and lowers the long-term cost of adopting and sustaining CSE methods in production.
    \item \textbf{Strengthen academic interactions:} Strengthen sustained interaction with universities and research groups, for example, through joint projects, shared benchmark cases, student exchange, and contributions to pre-competitive CSE software and data infrastructure. Continuous exchange accelerates the transfer of academic advances into practice while feeding real-world requirements back into research.
\end{itemize}

\noindent\textbf{What is the benefit for industry?}\\
These measures improve robustness, reduce duplication
of effort, and strengthen the long-term value of computational assets. They also
help translate academic developments into scalable products and operational
workflows while making resource consumption more transparent.

\subsection{Recommendations for Those with Project Responsibility}
\hphantom\\\vspace{-12pt}

\setlength{\intextsep}{0pt}
\begin{wrapfigure}{l}{20pt} 
\centering 
\vspace{-0pt}
{\fontsize{20pt}{30pt}\selectfont \textcolor{teal}{\faUsersCog }} 
\end{wrapfigure}
\noindent  Researchers with project responsibility lead funded initiatives, manage research teams, and oversee larger collaborative efforts and community codes. Their decisions shape how resources are used and how workflows are designed, and thus how sustainability principles are integrated into day‑to‑day research practice. Consequently, they play a pivotal role in steering computational science and engineering toward future developments. To advance sustainability in CSE, they should:

%JK
\begin{itemize}
    \item \textbf{Promote sustainability dimensions:} Promote methods and reporting strategies that relate scientific outcomes to computational effort and energy investment and help reduce the resources required. More broadly, keep environmental, social, and economic sustainability dimensions present during supervision and mentoring, so that the next generation adopts them as a natural part of good research practice.
    \item \textbf{Report measures taken:} Report transparently on the energy footprint of development and experiments, or on the measures taken to reduce it, as well as on the planned longevity of the developed code. Making these choices explicit at the project level normalizes energy awareness and signals that resource use is a legitimate dimension of research quality.
    \item \textbf{Ensure software visibility:} Choose publication and dissemination formats that give proper visibility to software, data, and negative findings about inefficient workflows. Surfacing these outputs lets others build on them and prevents hard-won lessons about what does not work from being lost.
    \item \textbf{Acknowledge software contributions:} Make sure all software contributions are attributed to their authors, both within the team and in external outputs. Consistent credit rewards the sustained maintenance work behind reusable code and makes such contributions visible in researchers' careers.
\end{itemize}

\noindent\textbf{What is the benefit for the CSE researcher?}\\
Sustainably reusable software accelerates knowledge generation, reduces duplication of effort, and increases the likelihood that others can build on the work. In turn, this improves scientific visibility, broadens impact, and can strengthen the researcher's own performance indicators.

\subsection{Recommendations for Teachers and Faculty}
\hphantom\\\vspace{-12pt}

\setlength{\intextsep}{0pt}
\begin{wrapfigure}{l}{20pt} 
\centering 
\vspace{-0pt}
{\fontsize{20pt}{30pt}\selectfont \textcolor{teal}{\faChalkboardTeacher }} 
\end{wrapfigure}
\noindent  CSE teachers and faculty define the intellectual foundations and training pathways that guide the long-term evolution of the field. Through curriculum design and classroom interaction from undergraduate to postgraduate levels, they influence how emerging CSE researchers understand sustainability, including the role of software and responsible resource use. By embedding these principles into the education of the next generation of scientists, they can be considered to have one of the most significant roles in making CSE sustainable as a field. Teachers and faculty should: 

% Mario
\begin{itemize}
    \item \textbf{Discuss energy consumption:} Integrate aspects of performance assessment and energy consumption into domain- or method-specific CSE lectures. Introducing these considerations where students already learn the underlying methods makes resource awareness a natural part of computational thinking rather than an afterthought.
    \item \textbf{Teach sustainable software development:} Integrate complementary aspects of research software engineering into domain- or method-specific CSE lectures. Exposing students early to testing, documentation, and reproducible workflows equips them to produce durable and reusable code throughout their careers.
    \item \textbf{Foster interdisciplinary exchange:} Foster communication across disciplinary boundaries so that sustainability challenges can be discussed in a shared language across methods, software, and applications. A common vocabulary helps students connect algorithmic choices to their computational cost and real-world impact.
    \item \textbf{Introduce in silico labs:} Offer introductory ``in silico'' laboratory courses, analogous to wet-lab courses, to equip students with good computational research practices from the outset. Hands-on practice turns abstract principles into concrete habits that students carry into their own research.
\end{itemize}

\noindent\textbf{What is the benefit for the CSE educators?}\\
Students are increasingly recognizing that these skills are relevant in both academia and industry. The courses that teach them become more attractive and also help educators recruit student assistants, thesis candidates, and future doctoral researchers who can contribute effectively to sustainable software and modeling projects.

\subsection{Recommendations for Young Scientists, Students, and Postdoctoral Researchers}
\hphantom\\\vspace{-12pt}

\setlength{\intextsep}{0pt}
\begin{wrapfigure}{l}{20pt} 
\centering 
\vspace{-0pt}
{\fontsize{20pt}{30pt}\selectfont \textcolor{teal}{\faUserFriends }} 
\end{wrapfigure}
\noindent CSE students and postdoctoral researchers form the core of the community that will be most directly affected by the recommendations outlined above. This younger cohort often drives momentum toward resource awareness and sustainable software practices, thereby influencing the future trajectory of the field. At the same time, their learning choices and day‑to‑day engagement with CSE methods shape how sustainability principles are adopted across both academic and industrial environments. We recommend that young scientists:

\begin{itemize}
    \item \textbf{Know the basics:} Learn the best practices for publishing and sharing software, including licenses, funding acknowledgments, and citation. Building literacy in research software engineering, performance analysis, and resource-aware computing early in your training pays off throughout your career.
    \item \textbf{Dare to write software:} Reusing established code saves effort, benefits from collective quality assurance, and lets you focus on the questions that are genuinely new. But research sometimes demands new paths, and not every idea fits an existing tool. When existing tools do not adequately support the research question, you should not hesitate to develop new software -- even from scratch.
    \item \textbf{Contribute software:} Contribute to persistent community code collections where appropriate, and insist on visibility for those contributions, or choose projects that offer individual recognition. Active participation strengthens shared infrastructure while making your own work more durable and citable.
    \item \textbf{Promote sustainable CSE:} Reach out, share, and celebrate your findings, including what did not work. By setting an example and engaging peers, young scientists influence and shape the practices of the next generation.
\end{itemize}

\noindent\textbf{What is the benefit for the young scientist?}\\
These practices improve reproducibility, collaboration, and employability. They also help young researchers make more visible and lasting contributions, whether they continue in academia, move into research infrastructure roles, or transition to industry.

\section{Outlook: Sustainable Computational Science and Engineering}
\label{sec:outlook}

%Making progress in this field calls for a concerted effort to increase our ability for complexity management in application-oriented utilization of CSE methods. We also need to push methodological innovation on high-dimensional optimization and multiscale and complex geometry CSE methods toward efficient and reliable utilization of modern CSE for the development of sustainable technologies with the potential to solve societal challenges, such as low-carbon footprint technologies, low-cost technological renewable energy solutions, and intelligent predictive maintenance and circular economy strategies.

The joint consideration of sustainable computing and sustainable software brings an underexplored methodological question into view: how can the resources consumed by CSE, the longevity of its software and knowledge, and the sustainability benefits enabled by its results be assessed jointly across their different timescales? Developing approaches to this question offers a natural opportunity for future CSE. In the future, progress in CSE will increasingly be measured not only by the accuracy, speed, or scalability of individual computations, but also by the degree to which computational resource investment is turned into scientific insight and information gain across complete workflows and life cycles. In other words, sustainable CSE is not only about reducing energy consumption or improving software quality, but also about
the degree to which methods, software, and workflows use resources responsibly, remain reusable over time, and support robust decision-making in sustainability-relevant contexts. Moving beyond isolated performance metrics toward such multi-objective, life-cycle-aware assessment creates a research opportunity for CSE: expertise in multi-objective optimization, optimal control, uncertainty quantification, and model hierarchies can help make trade-offs between immediate resource use and longer-term value explicit and tractable. A more sustainable CSE ecosystem will make advanced computational methods more trustworthy, reusable, and impactful across application domains. This, in turn, strengthens the ability of CSE to contribute meaningfully to societal challenges that require both technological innovation and responsible use of resources.

The examples discussed in this paper arise predominantly from model-based CSE and, more specifically, from simulation based on partial differential equations, reflecting the expertise represented at the workshop. They should be understood as illustrations rather than as delimiting the scope of CSE. The underlying considerations regarding resource-aware computation, sustainable software, and the relation between computational investment and scientific benefit also apply to other computational fields. Examining how these considerations translate across the broader methodological spectrum of CSE remains an important direction for future work.

However, substantial research remains necessary to realize this goal. The sustainability dimensions of CSE itself require further methodological, software-related, and institutional attention, even as this paper already formulates recommendations for different stakeholder groups. In particular, future research is needed on meaningful metrics, on the relation between computational effort and scientific benefit, and on ways to integrate sustainability considerations systematically into CSE workflows and software ecosystems.

One promising direction for such future work is to think of sustainable computing in CSE as a continuous cycle linking the measurement of resource consumption, the optimization of computational workflows across multiple objectives and timescales, the realization of corresponding practices in software and infrastructure, and their continued evaluation in light of evolving scientific, technical, and societal conditions. This perspective does not alter the goals of CSE, but helps to structure how sustainability-related progress may be made visible, assessed, and refined over time. In this sense, sustainable computing offers not only a set of immediate best practices but also a conceptual basis for long-term methodological and institutional development.

This need becomes even more urgent in light of the rapid development of generative AI and agentic systems. These systems are moving beyond assistance with isolated tasks toward the modification of repositories and the execution of extended software and computational workflows. They may accelerate software development and integration, but can also increase computational demand and make provenance, software dependencies, and individual contributions less visible. The sustainable-software practices emphasized in this paper therefore become more important, not less. At the same time, training and deploying increasingly capable foundation models comes with substantial computational and environmental costs. AI should therefore not only be used to advance sustainable CSE, but should also be developed and adopted according to the sustainability principles outlined in this paper.

Conversely, CSE has much to contribute to the sustainability of AI. Decades of research on exploiting physical principles, mathematical structure, adaptivity, multiscale modeling, and uncertainty quantification demonstrate that scientific insight does not necessarily require ever-increasing computational effort. Embedding such principles into AI methods—from physics-informed learning to structure-preserving architectures and hybrid simulation-AI workflows—offers a promising path toward AI systems that are not only more trustworthy and scientifically grounded, but also substantially more computationally efficient. In this way, sustainable CSE and sustainable AI should be viewed as mutually reinforcing research directions rather than independent developments.\\[3mm]

\section*{Acknowledgements}

The workshop underlying this paper was organized within the K\'arm\'an Conference Series program, which is generously sponsored by the Exploratory Research Space (ERS) of RWTH Aachen University. The ERS Series of K\'arm\'an Conferences features high-profile conferences on emerging research areas. These conferences aim to foster interdisciplinary collaboration at RWTH Aachen and position these areas within an outstanding research community.

\section*{Declarations}

\begin{itemize}
\item Funding: The workshop on which this paper is based was supported by the Exploratory Research Space (ERS) of RWTH Aachen University within the K\'arm\'an Conference Series.
\item Conflict of interest/Competing interests: The authors declare that they have no known competing interests.
\item Author contributions: The ideas presented in this manuscript were developed collaboratively before and during the workshop by all participating authors. The list of authors also shows who contributed to draft writing, reviewing, and editing the final text. The authors used AI tools, such as ChatGPT, Codex and RWTH GPT for language editing and quality-assurance support during manuscript preparation. These tools did not contribute scientific conclusions or original ideas, and the authors take full responsibility for the set of recommendations featured in this manuscript.
\item Data availability: No datasets were generated or analyzed specifically for this manuscript.
\item Code availability: No software was developed specifically for this manuscript. The paper discusses existing software projects referenced in the main text.
\item Supplementary information: No supplementary information is provided for this manuscript.
\end{itemize}

\bigskip


\bibliography{steinfeld}% common bib file

\end{document}